\documentclass{INTERSPEECH2023}
\usepackage{subfigure}

\interspeechcameraready

\title{Multi-Scale Temporal Transformer For Speech Emotion Recognition}
\name{Zhipeng Li$^1$, Xiaofen Xing$^1$\sthanks{{*} Corresponding author.}, Yuanbo Fang$^1$, Weibin Zhang$^3$, Hengsheng Fan$^4$, Xiangmin Xu\textsuperscript{1,2}}
\address{$^1$ South China University of Technology, Guangzhou, China 
\\ $^2$ Pazhou Lab, Guangzhou, China
\\ $^3$ VoiceAI Technologies, Shenzhen, China
\\ $^4$ Forensic Science Institute of Guangzhou Public Security Bureau}
\email{}

\begin{document}

\maketitle
 
\begin{abstract}
Speech emotion recognition plays a crucial role in human-machine interaction systems. Recently various optimized Transformers have been successfully applied to speech emotion recognition. However, the existing Transformer architectures focus more on global information and require large computation. On the other hand, abundant speech emotional representations exist locally on different parts of the input speech. To tackle these problems, we propose a \textbf{M}ulti-\textbf{S}cale \textbf{TR}ansfomer (MSTR) for speech emotion recognition. It comprises of three main components: (1) a multi-scale temporal feature operator, (2) a fractal self-attention module, and (3) a scale mixer module. These three components can effectively enhance the transformer’s ability to learn multi-scale local emotion representations. Experimental results demonstrate that the proposed MSTR model significantly outperforms a vanilla Transformer and other state-of-the-art methods across three speech emotion datasets: IEMOCAP, MELD and, CREMA-D. In addition, it can greatly reduce the computational cost.
\end{abstract}

\vspace{1em}
\section{Introduction}
In recent years, with the development of artificial intelligence and robotics, affective computing has become more and more important in human-computer interaction. Human emotions and intentions are well contained in the speech. Speech emotion recognition has a wide range of applications in spoken dialogue systems, call-center conversation analysis, etc. It can also be potentially used in a smart device. Despite great progress that has been made in speech processing, natural emotion understanding is still a challenging task for many smart systems.

With the advancement of deep neural networks, several attempts have been made to classify the emotional utterances through recurrent neural networks and convolutional neural networks. Guo et al. \cite{guo2021representation} proposed a CNN based on a spectro-temporal-channel attention module to improve emotion representation learning ability. 
Some algorithms\cite{li2019attentive,2020Speech,kumar2021towards} used RNN to model temporal sequence and attention mechanism for temporal tokens weighting have helped emotion representation extraction from speech.

Self-attention-based Transformers \cite{vaswani2017attention, devlin2018bert, dai2019transformer, dosovitskiy2020image} have become the main backbone in natural language processing(NLP) and computer vision(CV). Inspired by the success of NLP, researchers have tried to replace the entire CNN or RNN with a Transformer. Significant progress has been made in many speech-related tasks such as automatic speech recognition \cite{li2022transformer, xie22b_interspeech, audhkhasi22_interspeech}, speech enhancement \cite{dang2022dpt, deoliveira22_interspeech, kim21h_interspeech}. However, the computational resources that a Transformer with the full-attention mechanism is quadratic to the sequence duration, making it difficult to run on mobiles and embedded devices. In addition, its applications to the speech emotion recognition (SER) task remains limited since human emotions are inherently complex and ambiguous.
Some authors have proposed Transformer-based sparse attention mechanisms, such as BigBird\cite{zaheer2020big} in NLP, Image transformer\cite{parmar2018image}, Swin transformer\cite{liu2021swin} in CV, but these are not well suited for speech emotion tasks, since emotion is embedded in a long segment of continuous speech. So, it is necessary for us to design specific transformers for SER.
 Chen et al. \cite{chen2022key} proposed a Transformer based algorithm to capture all emotion features through a single fixed-scale size feature extractor. This might be inappropriate since human emotions can be expressed in different parts of the speech with different time spans. 
Zhu et al. \cite{zhu2022speech} used two different convolutional kernel sizes to simulate the extraction of multi-scale emotions from speech. We believe that using more different speech time scale information will be more helpful for emotion extraction. Emotional cues are multi-grained in nature, so a efficient light-weight Transformer that can utilize multiple granule levels of the acoustic features is more suitable for the speech emotion recognition (SER) task.

\begin{figure}[h]
\centering
\setlength{\abovecaptionskip}{0.cm}    
\includegraphics[width=8cm]{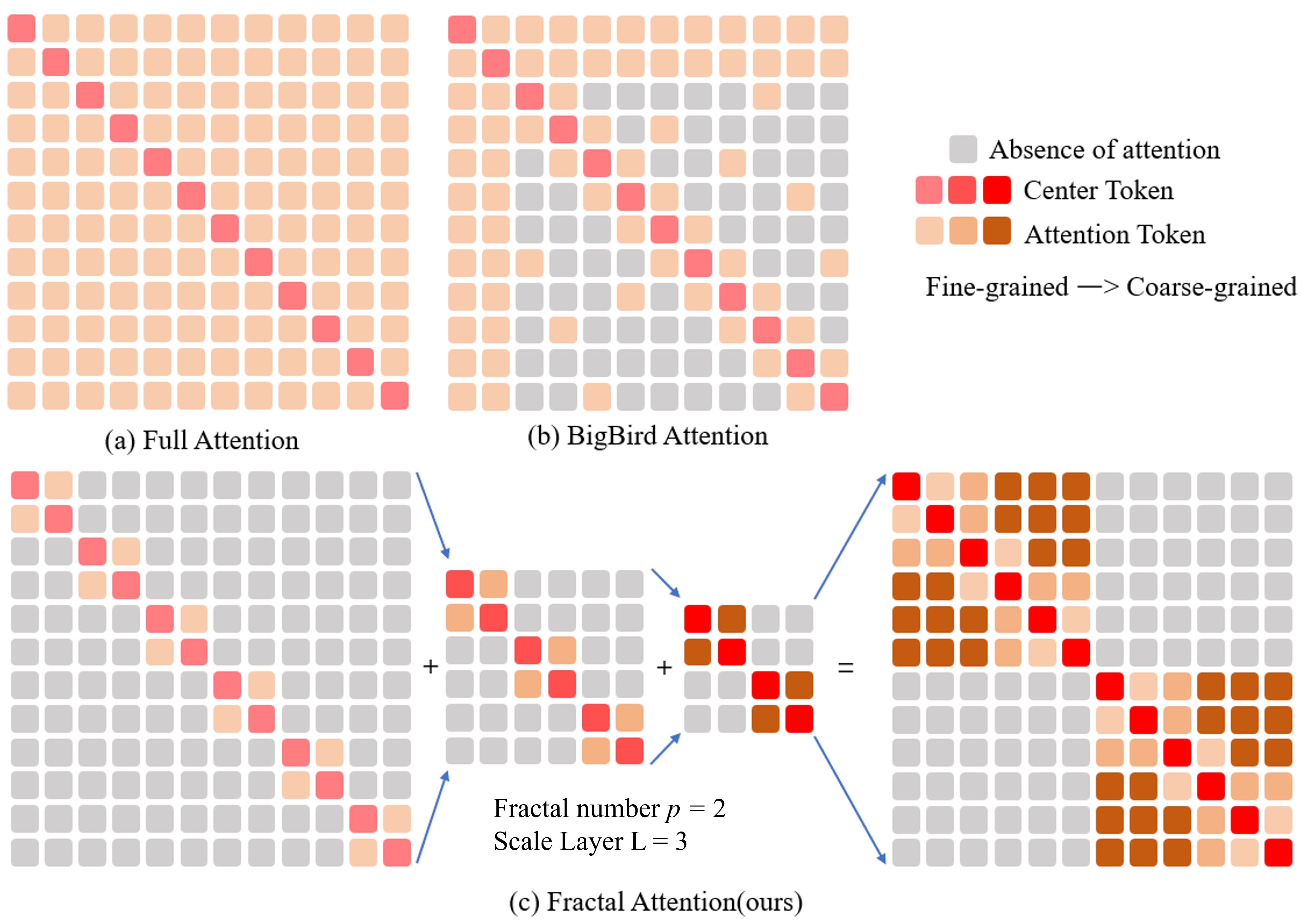}
    \caption{Compared with different attention mechanisms. Full attention means high computational cost and unnecessary attention redundancy and other sparse attention do not take well into account the characteristics of emotion in speech. We propose to extract emotional representation in different speech temporal scales.}
    \label{fig:galaxy}
\end{figure}

To deal with the limits of existing Transformers used in SER, we propose a  \textbf{M}ulti-\textbf{S}cale \textbf{TR}ansfomer(MSTR).

\begin{figure*}[ht]
    \centering
    \setlength{\abovecaptionskip}{0.cm}
    \includegraphics[width=17cm]{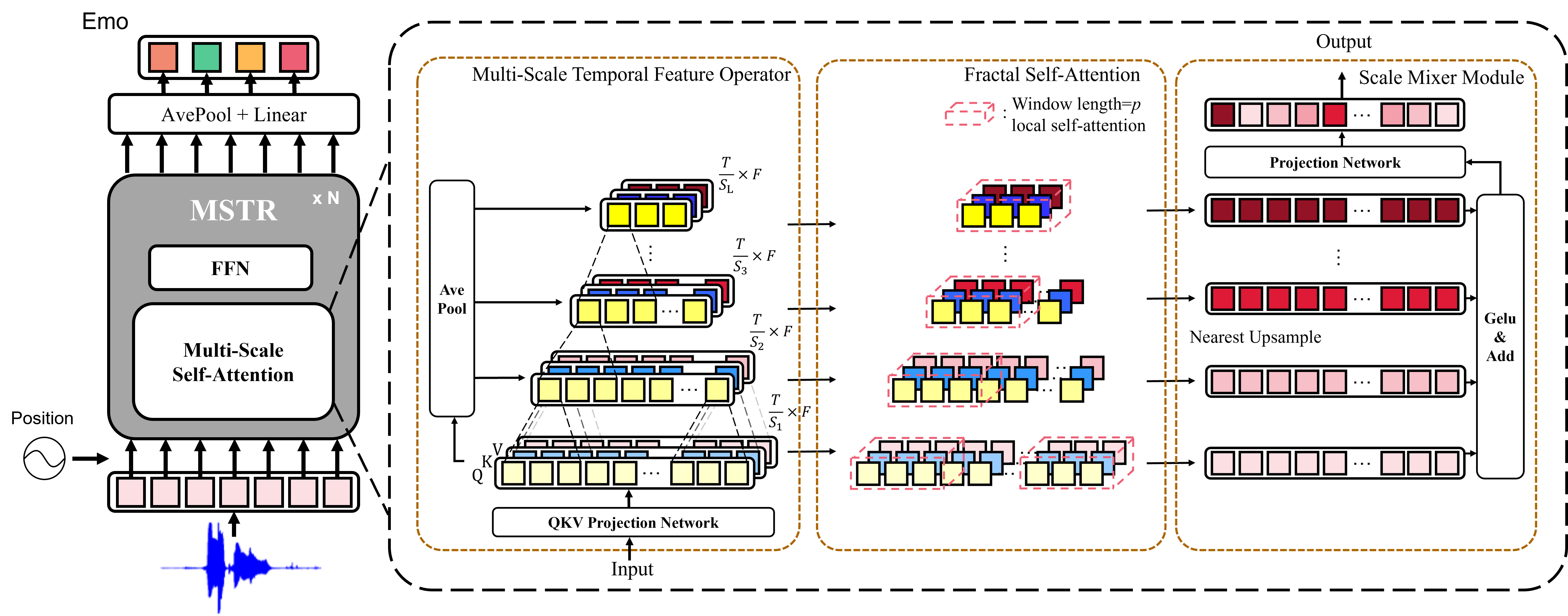}
    \caption{Overall architecture of the proposed Multi-Scale Transformer (MSTR) for speech emotion recognition. Each basic block of the MSTR mainly consists of three main components: a multi-scale temporal operator, a fractal self-attention module , and a scale mixer module. FFN means feedforward networks.}
    \label{fig:galaxy}
\vspace{1.0em}
\end{figure*}

The contributions of this article are summarized as follows:
\begin{itemize}
\item
To obtain multi-grained speech emotion representations, we proposed MSTR. The first effective transformer-based model that exploits multi-scale temporal features. Experiments demonstrate the great potential of the proposed method in modeling human emotion expressed in different temporal granularities in speech.
\end{itemize}

\begin{itemize}
\item
Experiments also show that the proposed model achieves comparable results with other state-of-the-art methods on both IEMOCAP\cite{busso2008iemocap}, Meld\cite{poria2018meld} and CREMA-D\cite{cao2014crema} SER benchmark datasets, but with much less computation requirement.
\end{itemize}

\section{Methodology}
The overall architecture of the proposed MSTR model is shown in Figure 2. As can be seen, different from a vanilla Transformer, the basic block of the proposed MSTR network mainly contains three components:  a multi-scale temporal feature operator, a fractal self-attention module, and a scale mixer module. The multi-scale temporal features operator takes raw acoustic feature or output from the lower layer as input and produces multiple output features with different temporal scales. The fractal self-attention module is used to efficiently model the temporal relations between different frames within a fixed-length window. Finally, the scale mixer module effectively fuses features at different temporal scales, to create a unified and mixed emotional feature representation. Compared to the original full attention mechanism, fractal attention is more effective in learning multi-grained features while greatly reducing model redundancy. Other modules like feed forward network remain the same as the original vanilla Transformer. The transformer's output will be fed to a classifier with three fully connected layers for sentiment classification. Details about these components are presented in the following sections.

\subsection{Multi-Scale Temporal Feature Operator}

We proposed a multi-scale temporal feature operator to parallelly extract multi-scale feature representations from raw acoustic features or output from lower layer. It takes a sequential feature \textbf{\textit{X}} $\in\mathbb{R}^{T \times F}$ as input, where \textit{T} is the number of input frames and  \textit{F} indicates the feature dimension. Similar to a vanilla Transformer, we first obtain the Query, Key, and Value. Specifically, the input \textbf{\textit{X}} is projected in to $Q=XW^{Q}$, $K=XW^{K}$, and $V=XW^{V}$, where $W^{Q}, W^{K}, W^{V} \in \mathbb{R}^{F \times F}$.

As shown in Figure 2, the obtained $\left\{Q, K, V\right \}$ are then fed into an average pooling module separately to get features at different time scales. Specifically, a scaling factor ${\textbf{S}}_{k}={p}^{k-1}$ is designed for the $k^{th}$ scale level, where \textit{p} is the fractal factor and \textit{k} $\in\left\{1,2, ...,L \right\}$. The $k^{th}$ scaling level operates on top of the ${k-1}^{th}$ level by averaging $p$ adjacent frames. In a word, the input feature set $\left\{Q, K, V\right \}$ goes through the average pooling operator to obtain temporal scale feature sets at different time scale $\textbf{\textit{X}}^{k}=\left\{Q^{k}, K^{k}, V^{k}\right\}\in\mathbb{R}^{\frac{T}{S_{k}} \times F}$. The new feature set $\textbf{\textit{X}}^{k}$ will be fed into the fractal self-attention module to model the features’ temporal relationship. It is worth mentioning that use of the pooling operator rather than the convolution can perform better, maintain the original timing structure, and wouldn't add any extra parameters.

\subsection{Fractal Self-Attention}
A vanilla Transformer uses global self-attention and thus requires large amount of computation. In the proposed MSTR model, we propose to calculate the self-attention within a fixed-length window since we already have features at different time scales. In our implementation, the length of the window is set to \textit{p}, i.e. same with the fractal factor. Thus we call the self-attention module the fractal self-attention.

Specifically, the feature set at the $k$-th scale level $X^{k}=\left\{Q^{k}, K^{k}, V^{k}\right\}$ is divided by the window size \textit{p}. Take $Q^{k}$ for example , we will have $Q^{k}=\left\{Q_{1}^{k}, Q_{2}^{k}, ..., Q_{w}^{k}\right\}$, where $w=\frac{T}{S_{k}\times p}=\frac{T}{p^{k}}$. For data in the $i$-th window block, self-attention is computed, i.e.
\begin{gather}
\vspace{1em}
\label{attention}
Attention(Q_i^k,K_i^k,V_i^k)=softmax(\frac{Q_i^k{K_i^k}^{t}}{\sqrt{F}})V_i^k
\vspace{0.5em}
\end{gather}
where t means transpose and $A_i^k=Attention(Q_i^k,K_i^k,V_i^k)$.

Finally, the output matrices from the self-attention computed at different windows are concatenated along the time dimension to produce new features $X^{k}\in\mathbb{R}^{\frac{T}{S_{k}}\times F}$:
\begin{gather}
    Y^{k}=Concat(A_{1}^k, A_{2}^k, ...,A_{w}^k)
\end{gather}
The output $Y^{k}$ from the $k$-th level scale will then be fed to the following scale mixer module. In Equation \ref{attention}, a single-head attention with full dimensionality is used for simplify. The multi-head mechanism described in \cite{vaswani2017attention} can also be applied straightforwardly. The window length \textit{p} applied to self-attention calculation help substantially reduce computation, especially when the length of the input data is large.

\subsection{Scale Mixer Module}
Fractal self-attention module can greatly reduce the computational complexity, but it also leads to the problem that the model might excessively focus on local semantics and ignore global information. We thus propose a scale mixer module to aggregate multi-scale data to get a unified emotion representation. The first step is to interpolate data at different time scales to have the original temporal sequence length \textit{T}. This is achieved by performing the nearest up-sample operation. Then the up-sampled data go through an activation function Gelu before we simply add all of them up. Finally, a linear projection $W^{O}$ is applied to get the final multi-scale emotion representation.
\begin{gather}
\setlength{\abovedisplayskip}{1pt}
\setlength{\belowdisplayskip}{1pt}
    Y_{up}^{k}=UpSampling(Y_{s}^{k}) \\
    Output=(\sum_{k=1}^N Gelu(Y_{up}^{k}))W^{O} 
\end{gather}
The local semantic information from fine-grained features and the global semantic information from coarse-grained features will be re-unified into a single representation. The scale mixer module effectively complements the deficiency of only being able to extract salient local information within a short window in a specific scale caused by the fractal self-attention. Other more complicated fusion methods like scale attention to select important emotional information may also be used to aggregate information from multi-scale features. But the simple method described above performs the best among all the fusion methods we explored in our experiments.

\subsection{Computational Complexity Analyse}
In a MSTR model, the self-attention is computed within a window of length \textit{p}. The computation complexities of the self-attention from a vanilla Transformer(VTR) and that from the fractal self-attention in a MSTR model are given below:
\begin{gather}
    \mathcal{O(VTR)}=T^{2}\times F \\
    \mathcal{O(MSTR)}=\sum_{k=1}^L{\frac{T}{S_{k}}}\times p^{2} \times F
\end{gather}
Here we ignore the computational effort of pooling and upsampling because they are much smaller than the computational effort of the self-attentive module. As can be seen, the computation for the self-attention layer in a vanilla Transformer is quadratic to the input sequence length, while the computation for the fractal self-attention in a MSTR model is liner. 

\section{Experiments}
\subsection{Datasets}
\textbf{IEMOCAP}: The dataset was used in the same way as in previous studies\cite{zou2022speech, chen2022speechformer}. The subset of IEMOCAP, which contains 5531 utterances of angry, happy(the category excited is labeled as happy), sad, and neutral was used. A 5-fold leave-one-session-out cross-validation strategy was utilized.

\noindent \textbf{Meld}: Meld is a multi-model dataset containing 13708 utterances with 7 emotion classes. There are 9989/1109/2610 utterances in the training/validation/testing sets and the performance on the test set are reported.

\noindent \textbf{CREMA-D}: It contains 7442 clips of 91 actors. All the clips are divided into six categories, i.e. neutral, happy, anger, disgust, fear, and sad. 80$\%$ the samples from CREMA-D were selected randomly as the training set and the remaining 20$\%$ as the test set, and the test set are reported.

\begin{table}[t]
	\centering
	\caption{Details about the three datasets used in the experiments. LR represents learning rate. Three common evaluation metrics, i.e. Weight Accuracy (WA), Unweighted Accuracy (UA), and Weighted average F1 (WF1) are used to access the model performance.}
	\label{tab:2} 
	\setlength{\tabcolsep}{1.8mm}
	\begin{tabular}{| c | c | c | c | c |}
		\hline Dataset & Feature & Epochs & LR & Evaluation Metric\\
		\hline IEMOCAP & Hubert & 150 & $1e^{-4}$ & WA $\&$ UA \\
		\hline Meld & Hubert & 150 & $5e^{-6}$ & WF1\\
		\hline CREMA-D & Hubert & 100 & $5e^{-5}$ & WA $\&$ UA\\
		\hline
	\end{tabular}
\end{table}

\subsection{Experimental Setup}
Details about the experimental settings as well as the evaluation metrics used for three datasets are listed in Table 1. We used a pre-trained Hubert(-large)\cite{hsu2021hubert}model to extract raw acoustic features. For both the MSTR model and the baseline model--vanilla Transformer, cross-entropy is employed as aloss function, and Adam \cite{kingma2014adam} optimizer is employed, the number of basic modeling blocks is 4, the number of heads is 16 and the batch size is 32. In our implementation, Fractal number \textit{p} is 3 and the number of scale layers \textit{L} is 4. In order to eliminate the effects of randomness, we trained and evaluated the models 10 times (setting 10 different seeds from 0 to 9), and report the average scores in the following.

\begin{table}[t]
	\centering
	\caption{Performance comparison of the proposed MSTR model, the baseline (a vanilla Transformer), and other
state-of-the-art methods on IEMOCAP, MELD, and CREMA-D. ``-'' means the original paper doesn't give the corresponding results. Params means number of model parameters and FLOPs means model computational complexity. The rules of calculating FLOPs we are consistent with vit\cite{dosovitskiy2020image,liu2021swin}}
	\label{tab:2} 
	\setlength{\tabcolsep}{1.8mm}

	\begin{tabular}{ c | c | c | c | c }
    	\hline \multicolumn{5}{c}{IEMOCAP}\\
		\hline  Method & Params & FLOPs & WA & UA\\
		\hline
		Guo (2021)\cite{guo2021representation} & - & - & 61.32 & 60.43\\
		Wang (2021)\cite{wang2021learning} & - & - & 66.50 & 65.70\\
		Chen (2022)\cite{chen2022speechformer} & 16.72M & 2.28G & 62.90 & 64.50\\
		Gudmalwar (2022)\cite{gudmalwar2022magnitude} & - & - & - & 67.42\\
            Fan (2020)\cite{fan2022isnet} & - & - & 70.40 & 65.00\\
		Zou (2022)\cite{zou2022speech} & - & - & 69.80 & 71.05\\
		Baseline & 27.0M & 892.2M & 68.90 & 70.02\\
		\textbf{MSTR(ours)} & 27.0M & \textbf{33.5M} & \textbf{70.60} & \textbf{71.60}\\
            \hline
    	\hline \multicolumn{5}{c}{Meld}\\
    	\hline  Method & Params & FLOPs & \multicolumn{2}{|c}{WF1}\\
		\hline Liang (2020)\cite{liang2020semi} & - & - & \multicolumn{2}{|c}{40.20}\\
		Lian (2021)\cite{lian2021ctnet} & - & - & \multicolumn{2}{|c}{38.20}\\
		Chen (2022)\cite{chen2022speechformer} & 33.2M & 1.64G & \multicolumn{2}{|c}{41.90}\\
		Hu (2022)\cite{hu2022mm} & - & - & \multicolumn{2}{|c}{42.72}\\
		Baseline & 25.3M & 432.7M & \multicolumn{2}{|c}{45.20}\\
		\textbf{MSTR(ours)} & 25.3M & \textbf{29.8M} & \multicolumn{2}{|c}{\textbf{46.15}}\\
		\hline
    	\hline \multicolumn{5}{c}{CREMA-D}\\
		\hline  Method & Params & FLOPs & WA & UA\\
		\hline
		Baseline & 24.1M & 205.9M & 77.70 & 77.90\\
		\textbf{MSTR(ours)} & 24.1M & \textbf{27.16M} & \textbf{79.70} & \textbf{79.92}\\
		\hline
	\end{tabular}
    \vspace{-1em}
\end{table}

\subsection{Result and Discussion}
\subsubsection{Comparison Analyse}
The results of different methods are shown, in Table 2. As can be seen, compared with the baseline model (a vanilla Transformer), the proposed MSTR significantly improves the performance on all the results: +1.70$\%$ WA and +1.58$\%$ UA in IEMOCAP, +0.95$\%$ WF1 in Meld, +2.00$\%$ WA and +1.98$\%$ UA in CREMA-D. In addition, it reduces FLOPs up to 96.25$\%$ in IEMOCAP, 93.11$\%$ in Meld, 86.81$\%$ in CREMA-D. The proposed method uses local self-attention within a window to model the input’s correlation rather than full attention. On the other hand, it expands the local receptive field through a multi-scale structure. The results demonstrate the effectiveness of the proposed MSTR model.

The MSTR model also outperforms some well-known systems on the three corpora. On IEMOCAP, MSTR achieves comparable performances to \cite{zou2022speech}: +0.80$\%$WA and +0.55$\%$UA. In terms of computation, it only requires about 1.5$\%$ of the computation of the model in \cite{chen2022speechformer}. On Meld, the MSTR model outperforms the previous methods by a large margin: +3.43$\%$WF1 over \cite{hu2022mm}. In summary, the MSTR architecture achieves state-of-the-art performance and computational efficiency in three popular benchmark datasets in speech emotion recognition.

\subsubsection{Significant Hyperparameter Analysis}
We also did experiments to evaluate the influence of the two important hyperparameters: the fractal factor \textit{p} and the number of scale layers \textbf{\textit{L}}. Figure 3 shows the results compared with baseline model. As can be seen from Figure 3a, setting \textit{p} to 3 achieves the best performance in all three speech emotion datasets. From Figure 3b, we can clearly see that the model can substantially benefit from the multi-scale configurations. The performance of the MSTR model drops on all the corpus when the number of scale levels goes from 4 to 1, which demonstrates the effectiveness of multi-scale temporal transformer. When \textit{p} set 3 and \textbf{\textit{L}} set 1, the fractal attention mechanism degenerates to general window-based attention, compared with baseline model, MSTR does not perform well, and that means the window-based attention mechanism does not apply to all speech emotion datasets, and multi-scale attention mechanism can achieve great performance with low computing volume .This in turn confirms that human emotion indeed exists in features with different time-scales and the fact that multi-grained emotion representations are essential. 
Rational use of emotion representations implied by speech in different time scales is the key to the speech emotion recognition task.

\begin{figure}[t]
    \centering
    \subfigure{\includegraphics[width=8cm]{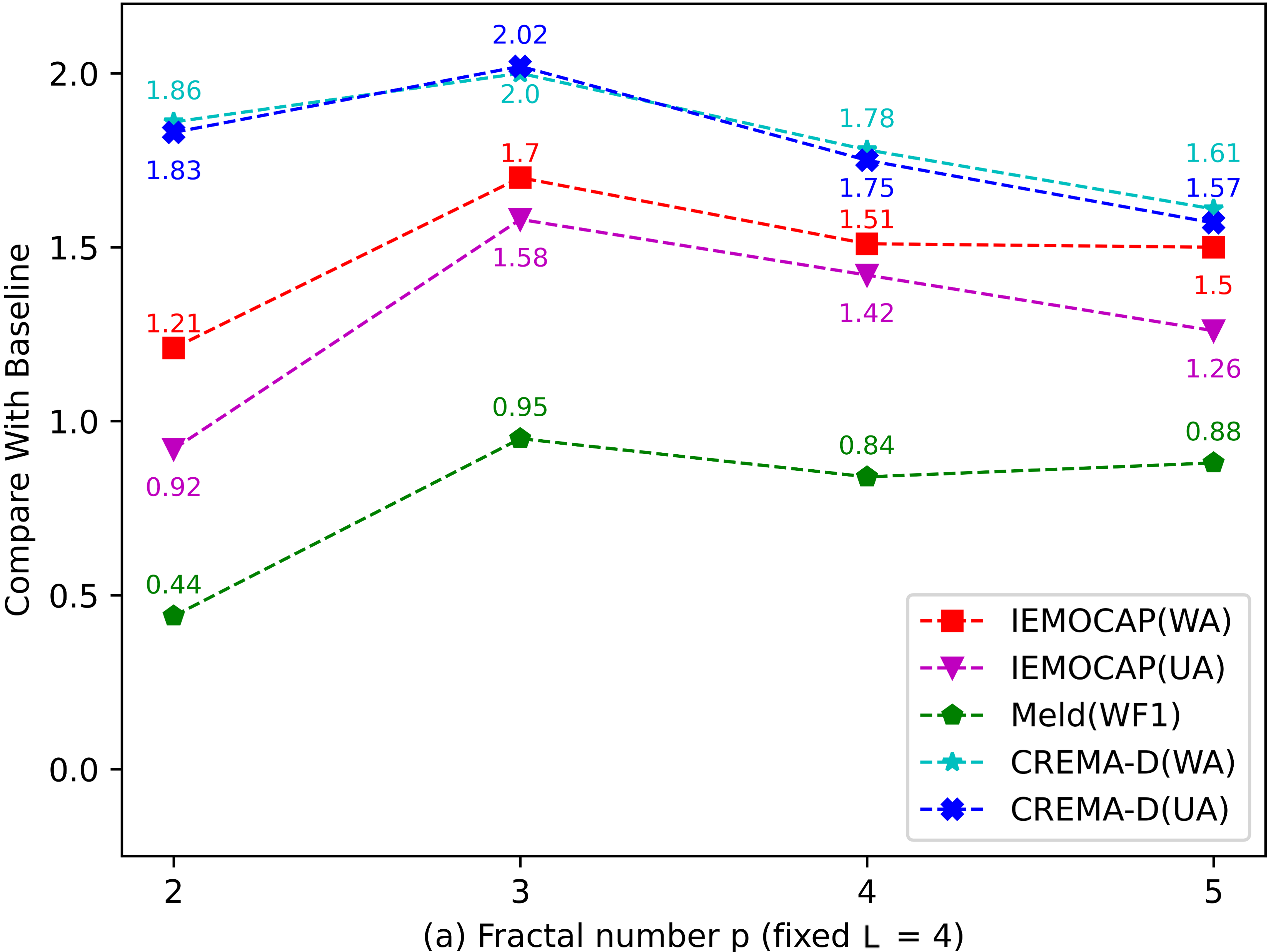}}
    \subfigure{\includegraphics[width=8cm]{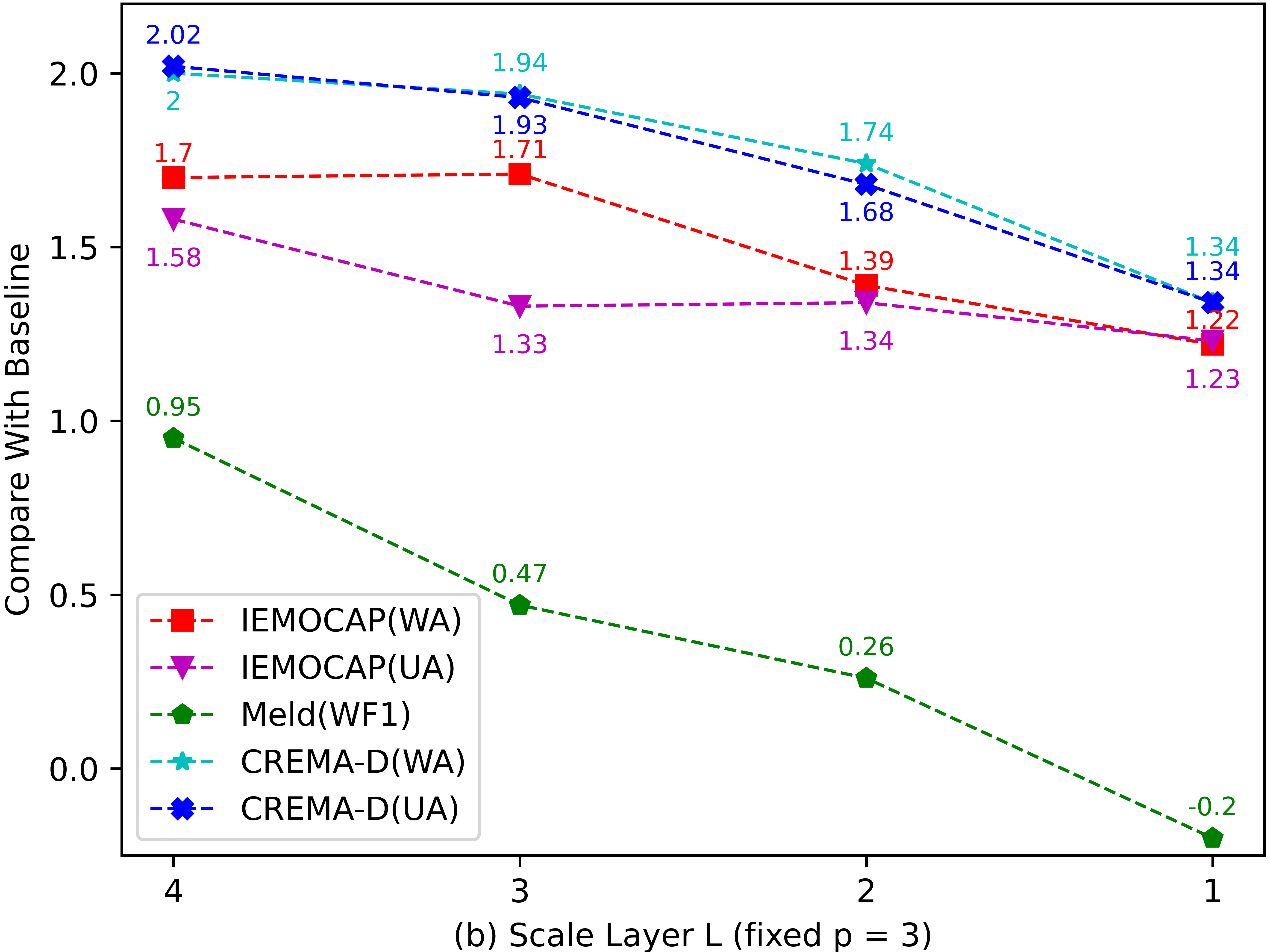}}

    \caption{Analysis of the influence of hyperparameters \textit{p} and \textbf{\textit{L}}}
\end{figure}
\vspace{1.0em}
\section{Conclusions}

In this paper, we delve into the emotional multi-scale representation learning and propose a new efficient architecture named MSTR for the speech emotion recognition task. The fractal self-attention module in MSTR can effectively extract multi-scale temporal emotion features. The scale mixer module summarizes multi-scale representations into a single emotion representation. The ablation experiments show that multi-scale features contain more salient emotional information than single-scale features. MSTR was evaluated on three SER datasets. It outperforms the latest Transformer-based network in both computational complexity and performance. The MSTR model achieves new state-of-the-art results. 

\vspace{-0.5em}
\section{Acknowledgements}

The work is supported in part by the Science and Technology Project of Guangzhou 202103010002; 
in part by Natural Science Foundation of Guangdong Province 2022A1515011588; in part by Nansha key project 2022ZD011; 
in part by the National Natural Science Foundation of China U1801262; 
in part by the Fundamental Research Funds for the Central Universities 2022ZYGXZR0075; 
in part by Key-Area Research and Development Program of Guangdong Province 2022B0101010003; 
in part by the Guangdong Provincial Key Laboratory of Human Digital Twin 2022B1212010004.

\clearpage
\bibliographystyle{IEEEtran}
\bibliography{mybib}

\end{document}